\documentclass[prl,twocolumn]{revtex4}%
\usepackage{amsmath}
\usepackage{amsfonts}
\usepackage{amssymb}
\usepackage{graphicx}%

\begin{document}
\title{Extremely Sub-wavelength Planar Magnetic Metamaterials}
\author{W. - C. Chen, C. M. Bingham, K. M. Mak, N. W. Caira, and W. J. Padilla}
\email{Willie.Padilla@bc.edu}

\address{Department of Physics, Boston College, 140 Commonwealth Ave., Chestnut Hill, MA 02467, USA}

\begin{abstract}
We present highly sub-wavelength magnetic metamaterials designed for operation at radio frequencies (RFs). A dual layer design consisting of independent planar spiral elements enables experimental demonstration of a unit cell size $(a)$ that is $\sim$700 times smaller than the resonant wavelength $(\lambda_0)$. Simulations indicate that utilization of a conductive via to connect spiral layers permits further optimization and we achieve a unit cell that is $\lambda_0/a \sim 2000 $. Magnetic metamaterials are characterized by a novel time domain method which permits determination of the complex magnetic response. Numerical simulations are performed to support experimental data and we find excellent agreement. These new designs make metamaterial low frequency experimental investigations practical and suggest their use for study of magneto-inductive waves, levitation, and further enable potential RF applications.
\end{abstract}
\maketitle

Experimental verification of a negative index of refraction\cite{Smith00,Shelby01} confirmed a four decade old theoretical prediction\cite{Veselago68} and initiated a new area of research in artificial electromagnetic (EM) materials\cite{Pendry99}, known as metamaterials (MMs). In the last decade metamaterials research has yielded many exotic effects including: invisibility cloaking\cite{cloak}, perfect lensing\cite{Pendry00,superlens}, and perfect absorbers\cite{blackbody}. Interest in metamaterials stems from their available use in nearly any band of the electromagnetic spectrum and their ability to achieve almost any desired response\cite{MaM1,MaM2,MaM3}.

While there are relatively few technical restrictions limiting operational frequencies of metamaterials there are, however, some practical hurdles. To date there has been a noticeable lack of metamaterials demonstrated at low frequencies -- on the order of 100 MHz or lower. Although metamaterials may still be of use in this frequency range, the main drawback is that the dimensions can become impractically large. Wavelengths for radio frequencies range from thousands of meters (for 100's kHz) to lengths of meters (100's MHz). Since it is not uncommon for individual metamaterial elements to have sizes on the order of $\sim \lambda_0 /10-\lambda_0 /100$, (where $\lambda_0$ is free space wavelength), the side length of a single RF metamaterial element could be expected be around 100 m to 10 cm. In many situations this size may be unacceptable for real-world applications, particularly so for `portable' devices. The development of a new class of metamaterials with a significant reduction in the ratio of its physical size `$a$' to its resonant wavelength $\lambda_0$ may create new opportunities for low frequency metamaterial-based applications\cite{Magneto02,Wiltshire07,RFM1,RFIDM1}.

Magnetic properties exhibited by natural materials are typically weaker and less common than their electric counterpart. This is due to the fundamental difference in how magnetic and electric responses are generated - the former being produced either from orbital currents or intrinsic unpaired spins. This fundamental lack of diversity in materials hinders full development of electromagnetic devices. Metamaterials with a $\mu$-negative response have been shown to improve the effectiveness of systems that make use of magnetic fields at radio frequencies such as increasing the resolution of MRI\cite{Wiltshire03,Marques05} and enhancing wireless power transfer efficiency\cite{WPT1,WPT2}.

Here we propose that dual layer metamaterial spirals may be used in order to achieve extremely sub-wavelength RF magnetic metamaterials. We utilize and detail a time domain method for characterization of the complex magnetic response $(\mu(\omega))$ of the RF metamaterials and support experimental results with simulation.

As a starting point for exploration of RF magnetic metamaterials we begin with a design based on a planar spiral structure capable of achieving large inductances\cite{Marques04,mini07,Super10} -- as it is trivial to wind a significant length of wire into a small area. The spiral response can be approximated based on a simple $LC$ resonator model $\omega_{0}=1/\sqrt{LC}$, where $\omega_{0}$ is the resonant angular frequency, $L$ is the inductance, and $C$ is the capacitance of the MM. Additional reduction of the unit cell size may thus be achieved by maximizing $L$ and $C$ simultaneously. In single layer planar spirals the capacitance arises solely due to interactions between adjacent metallic windings\cite{Marques03}. However, since typical lithographic methods produce extremely thin metallic layers the capacitive values obtained in this manner are small. We may drastically increase the capacitance by adding a second spiral layer, thus creating an interlayer interaction between the two planar structures. In this way both the larger area of the broadside metallic surfaces and a supporting dielectric in-between will provide an increased capacitance. A similar idea had been demonstrated in the microwave range, ``the broadside-coupled'' split ring resonators (SRRs) were found to have a resonance occurring at much lower frequency than the original single layer design\cite{Marques02,Marques03}. The addition of the second layer will also provide an enhancement of the metamaterial inductance if added in series.

\begin{figure}
\begin{center}
\includegraphics[width=3.2in,keepaspectratio=true]{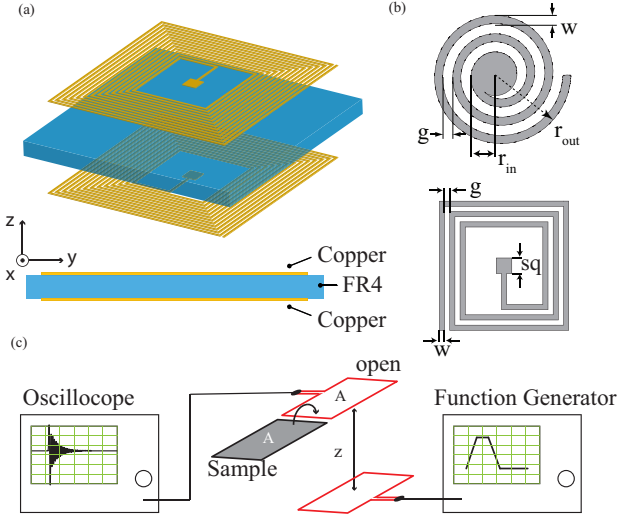}%
\caption{(Color online) (a) Schematic (exploded view) of the dual-layer square spiral metamaterial (top) and a side view showing material layers (bottom). (b) Schematic drawings of circular (top) and square spirals (bottom). (c) Schematic of the experimental apparatus as described in the text.}
\label{Fig1}%
\end{center}
\end{figure}

We investigate two different dual layer planar spirals - circular and square windings, see Fig. \ref{Fig1} (b). Simulations were performed with a commercially available 3D full wave finite element frequency domain solver. Excitation from a waveguide port is used where the electric field is along the $\hat{x}$ direction, and the magnetic field is along $\hat{z}$ - see Fig. 1(a). The simulated circular spiral design has: 21 turns, a linewidth $w$ = 170 $\mu$m, line spacing $g$ = 320 $\mu$m, inner radius $r_{in}$ = 1.2 mm, outer radius $r_{out}$ = 11.7 mm, and unit cell size (\emph{a}) of 25.4 mm. The square spiral has: 13 turns, a linewidth $w$ = 210 $\mu$m, line spacing $g$ = 320 $\mu$m, side length of the central capacitive pad $sq$ = 1.9 mm, and the side length of the square coil is 25 mm with a unit cell size (\emph{a}) of 26 mm. The dielectric slab for both patterns is 203 $\mu$m thick and was modeled using a dielectric value of $3.6+0.13i$. All metallic components were modeled as copper with a conductivity of $\sigma=5.8 \times 10^{7}$ S/m. For all cases, a vacuum box ($\emph{a} \times \emph{a} \times 40 (mm)$) was used, and the geometry of the spiral, including the size of the centrally located capacitive pad, was modified in order to obtain the greatest oscillator strengths. The simulated scattering parameters, $S_{11}$ and $S_{21}$, allow one to extract effective parameters via the transfer matrix method approach\cite{Smith05,AM}.

\begin{figure}
\begin{center}
\includegraphics[width=3.2in,keepaspectratio=true]{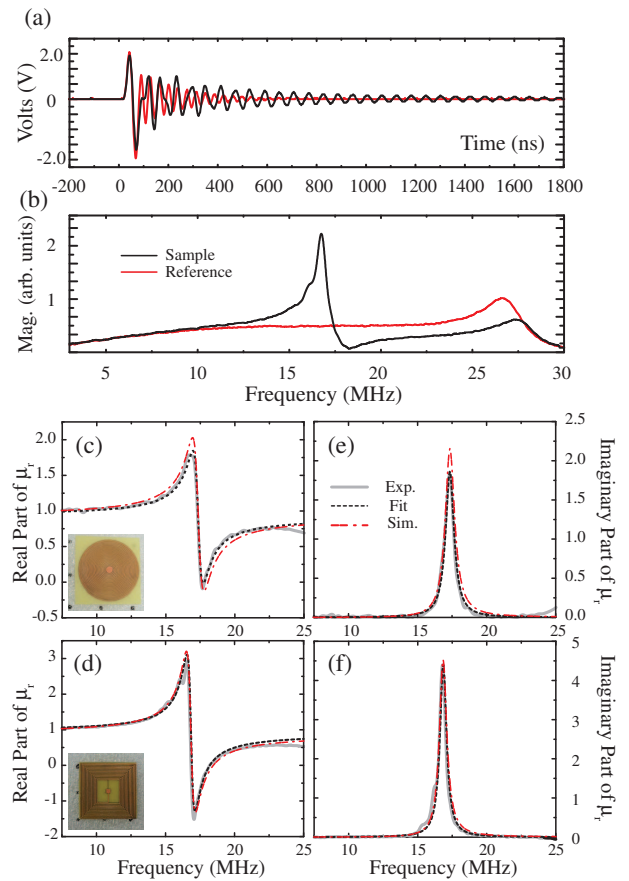}%
\caption{(Color online) (a) Time domain waveform recorded by the pickup coil for the metamaterial (black curve) and a reference signal (red curve). (b) The amplitude spectrum for the metamaterial (black) and reference (red) obtained by a Fourier transformation of (a). Experimental (grey curves), simulated (red dashed curves) and fits (black dotted curves) to the real (c,d) and imaginary (e,f) parts of $\mu_r$. Panels (c) and (e) show the real and imaginary permeability, respectively, for the circular spiral. Panels (d) and (f) show the real and imaginary permeability, respectively, for the square spiral.}
\label{Fig2}%
\end{center}
\end{figure}

The extracted permeability for each magnetic metamaterial is shown as red dash curves in Fig. \ref{Fig2} (c,e) for the circular spirals and (d,f) for the square spirals. A prominent magnetic Lorentzian type resonance can be observed for each configuration described by,

\begin{equation}
\tilde{\mu}_{r}(\omega) = \mu_{\infty} + \frac{\omega_{mp}^{2}}{\omega_{0}^{2}-\omega^{2}-i\gamma\omega}
\label{lorentzian}
\end{equation}

\noindent where $\mu_{\infty}$ is the relative permeability at frequencies higher than the resonance, $\omega_{mp}$ is the magnetic plasma frequency, $\omega_{0}$ is the center frequency of the resonance, and $\gamma$ is the loss.

In order to characterize the complex permeability $(\widetilde{\mu}(\omega)=\mu_1+i\mu_2)$ response of our magnetic metamaterials, ($\mu_1$ and $\mu_2$ are the real and imaginary portions of the magnetic response function, respectively), we utilized a time-domain spectroscopic method. A schematic illustrating our experimental apparatus is shown in Fig. \ref{Fig1} (c). In our setup a coil is driven by a voltage pulse from a function generator creating a transient magnetic field. A second coil (pickup coil) is co-axially aligned with the drive coil, (spaced a distance $z$ away) and measures the induced time dependent electromotive force $(emf(t))$ due to the incident magnetic field created by the drive coil. Based on Faraday's law of induction, the induced $emf$ of the reference and the samples can be expressed as,

\begin{equation}
\begin{array}{lcr}
emf(t)_r = -A\mu_0 \frac{dH_z}{dt} \\
emf(t)_s = -A\mu_r \mu_0 \frac{dH_z}{dt}
\end{array}
\label{emf}
\end{equation}

\noindent where $A$ is the area of the pickup coil, $\mu_0$ is the free space permeability, and $H_z$ is the $\hat{z}$ component of H field.

The effective permeability ($\mu_r$) of the samples can then be trivially derived from Eq. \ref{emf}, when the areas $A$ of the pickup coil and metamaterial samples are made equal.

\begin{equation}
\widetilde{\mu}_r(\omega) = \frac{\hat{F}(emf(t)_{s})}{\hat{F}(emf(t)_{r})}
\label{mu}
\end{equation}

In Eq. (3) $\hat{F}$ stands for the complex Fourier transform.

The statement that the effective permeability can be described by Eq. 3 above carries with it the assumption that the time dependence of the $\mathbf{H}$ fields between the sample and reference cancel. Although this is a reasonable supposition we next show that this follows directly from Maxwell's equations and is verified by the experimental results.

Ampere's law specifies that the bound current, which is related to the magnetization via $\mathbf{\bar{J}}_B=\mathbf{\bar{\nabla}} \times \mathbf{\bar{M}}$, and thus to the magnetic permeability, is what changes upon placing a magnetic material in the loop of the pickup coil. As $\mathbf{H}$ is only related to the true free current in Maxwell's Equations, this remains constant in our measurements and thus cancels for both the sample and reference measurements. Of course the above assumes that there is no change in the mutual coupling between the drive and pickup coils due to placement of the measured material, which would change the free current. We use the time domain method described here to characterize the complex magnetic response of metamaterials. In all cases we have made the areas of the pickup coil and metamaterial samples equal.

Metamaterial samples were fabricated by using standard printed circuit board (PCB) photolithography. All fabricated dimensions were identical to those simulated, and photographs are shown as the insets to Fig. \ref{Fig2}. We use the time domain method described above to characterize the complex magnetic response of metamaterials, see Fig. \ref{Fig1} (c). The time dependent magnetic field was generated using a coil antenna driven by a transient voltage. The peak to peak voltage ($V_{pp}$) was set to 10 V for a period of 80ns with a duty cycle of 37.5\%. A receiving coil spaced a distance of 8 cm away was connected to a digital sampling oscilloscope which captured the time dependent electromotive force induced by the incident magnetic flux. The shape and dimensions of the receiving coil were chosen such that it matched that of a 3 $\times$ 3 array of our magnetic metamaterials. For each sample we performed two measurements. The first was a reference where nothing was placed inside the receive coil. For sample characterization we placed the metamaterial array into the center of the receiving coil. All measurement were collected with an average of 128 wave forms.

Representative time pulses are shown in Fig. \ref{Fig2} (a) for both the sample (black curve) and reference measurement (red curve). A noticeable difference can be seen between the two measured time pulses as the reference signal has almost completely died out after 600ns, while the metamaterial signal `rings' for a much longer time (over 1800ns). A plot of the frequency spectra calculated from the Fourier transform of the time data is shown in Fig. \ref{Fig2} (b). Applying the extraction method of Eq. \ref{mu}, outlined earlier, one can determine the effective permeability $\mu_{r}(\omega)$ of the magnetic metamaterials.

The real and imaginary portions of the measured permeability are plotted as grey curves in Fig. \ref{Fig2} for both the circular (top panels) and square (bottom panels) samples. As can be observed both spiral metamaterial structures yield magnetic resonances with Lorentz-like lineshapes. The circular spiral has its resonance centered at 17.3 MHz and the square spiral at 16.8 MHz. The real part of the measured relative permeability for both designs reaches a negative value of -0.25 for the circular design and -1.51 for the square design. The full width at half maximum (FWHM) is 0.554 MHz and 0.794 MHz for the circular and square designs, respectively. The calculated Q-factors are 30.3 and 21.8 and both designs achieve decent oscillator strength at resonance. The simulated permeabilities (red dashed curves in Fig. \ref{Fig2}(c)-(f)) are nearly identical to the measured ones. For example the simulated and experimental peak $\mu$ values differ by only 0.2\% and 0.1\% for the square and circular spirals, respectively. Some discrepancies in the amplitudes and center frequencies for both designs are observed. Nevertheless, the experiments are in excellent agreement with the simulations. The errors can be attributed to imperfections from the fabrication process.

We compared experimental permeabilities to a frequency-dependent Drude-Lorentz oscillator, as shown in Eq. \ref{lorentzian}. The black dotted curves of Fig. \ref{Fig2} are fits to the experimental data and we find good agreement with parameters of: $\mu_{\infty} = 0.893$, $\omega_{mp} = 2\pi \times 4.91$ (MHz), $\gamma = 2\pi \times 0.745$ (MHz), and $\omega_{0} = 2\pi \times 17.3$ (MHz) for the circular geometry, and  $\mu_{\infty} = 0.87$, $\omega_{mp} = 2\pi \times 6.6$  (MHz), $\gamma = 2\pi \times 0.587$ (MHz), and $\omega_{0} = 2\pi \times 16.8$ (MHz) for the square geometry.

Radio frequency magnetic metamaterials investigated here yield reasonably strong resonances although their physical size $(a)$ is only a small fraction of their resonant wavelength. Circular spiral metamaterials achieved $a=\lambda_0 /683$ and square designs $a=\lambda_0 /703$. In comparison to simulated results for a single layer spiral, (not shown), the resonant wavelengths of the dual layer designs are nearly six times smaller, although all have roughly the same unit cell size. It's worth noting that in-spite of the strong sub-wavelength sizes of metamaterials presented here, there are simple improvements which will further increase $\lambda_0/a$. For example we have chosen to fabricate designs in which the two planar spirals are independent and only capacitively coupled through the substrate. We chose this geometry simply because these structures are easier to fabricate. However by connecting the two layers with a conductive via one may push sizes of these structures even more sub-wavelength, as the unwound length effectively doubles.

\begin{figure}
\begin{center}
\includegraphics[width=3.5in,keepaspectratio=true]{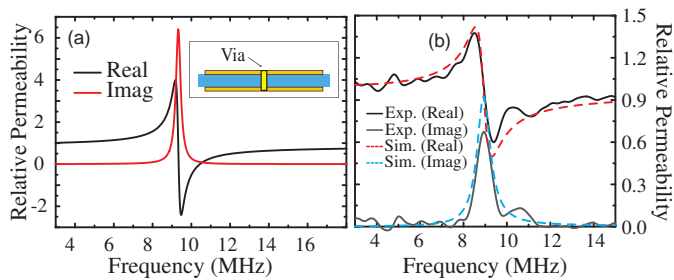}%
\caption{(Color online) (a) Simulated real (black) and imaginary (red) relative permeability for a dual layer circular spiral metamaterial with an interlayer via. (b) Experimental (solid curves) and simulated (dashed curves) relative permeability of a circular spiral metamaterial with conductive via - described in the text. }
\label{Fig3}%
\end{center}
\end{figure}

Figure \ref{Fig3} presents simulations of the square spiral, with a connecting via, and a unit cell size $a$ = 15.5 mm, linewidth $w$ = 100 $\mu$m, line spacing $g$ = 100 $\mu$m, board thickness of 203 $\mu$m, and consisting of 25 windings. Simulations indicate this design should exhibit a strong magnetic resonance at 9.34 MHz and the extracted complex $\mu$ is shown in Fig. \ref{Fig3} (a). For this geometry we find $a=\lambda_0 /2072$. Although the design discussed above is well within both laboratory and commercial fabricational capabilities, here we chose to experimentally demonstrate a slightly less subwavelength structure, and simply modify the circular spiral shown in Fig. \ref{Fig1} (a) and (c). Holes were drilled through the center capacitive pad of each unit cell and a wire was soldered on both sides to create the connecting via. We then characterized the sample and extracted the complex permeability as shown in Fig. \ref{Fig3} (b). We find a resonant frequency of 8.94MHz and thus this modified spiral magnetic metamaterial yields $a=\lambda_0/1321$. It should be noted that the experimental structure presented here was by no means an optimized design, i.e. we did not fine tune capacitive and inductive reactances which could be responsible for the significantly lowered oscillator strength. Additionally a notable Ohmic loss resulted from the manner in which the conductive via was experimentally implemented. The metamaterial geometry was simulated where the via was modeled as a resistive element (R=13$\Omega$) and we find excellent agreement with experiment as shown in Fig. \ref{Fig3} (b).

\begin{figure}
\begin{center}
\includegraphics[width=3.5in,keepaspectratio=true]{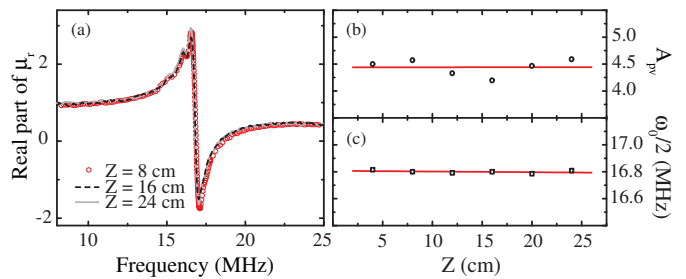}%
\caption{(Color online) Simulated results: (a) Measured permeability for several values of coil separation $z$. The peak to valley amplitude ($A_{pv}$) of $\mu_{r}$ (b) and the resonant frequency ($\omega_{0}/2\pi$) (c) versus the coil separation distance (z). Solids red lines are linear fits to the data.}
\label{Fig4}%
\end{center}
\end{figure}

As a verification to the validity of the experimental time-domain method presented here, we performed measurements of the response of magnetic metamaterials for several coil separation distances ranging from z = 4 cm to z = 24 cm. In all cases the sample under investigation was placed directly in the plane of the pick-up coil. As before all sample measurements were characterized and compared with a reference measurement. The induced maximum peak-to-peak value of the $emf$ for the reference measurements was observed to change from 2V for the closest spacing to 40 mV for the farthest. Calculated $\mu_r$ values are shown for three characteristic separations of z=8, 16, and 24 cm in Fig. \ref{Fig4} (a). In Fig. \ref{Fig4} (b) and (c) we plot, respectively, the peak-to-valley amplitude of $\mu_r(\omega)$ and the resonance frequency $\omega_0$, both as a function of coil separation. The red horizontal lines of Fig. \ref{Fig4} (b,c) are fits to the data and have nearly zero slope to within 10$^{-4}$. The measured center frequencies only deviate by a maximum of 1\% of the average value for $z$ distances studied here. Variation, however, in the peak-to-valley $\mu_r$ values is as much as 5\% though we attribute this discrepancy to noise in the system.

We computationally and experimentally demonstrated extremely sub-wavelength planar magnetic metamaterials. Experimental resonant wavelength to unit cell size values were shown to be as much as $\sim$1300, and an ideal simulated design was shown to be $\lambda_0/a\sim2000$. We find excellent agreement between experimental and simulated magnetic permeabilities, for all structures investigated, and attain reasonable oscillator strengths. A novel characterization method has been presented capable of determining the complex magnetic response of planar materials at RFs. Compact sub-wavelength designs presented here make low-frequency study of metamaterial practical which may lead to RF metamaterial applications.

We acknowledge support from the Department of Energy under grant DE-SC0002554. The authors would also like to thank the D.~R. Smith and Y.~A. Urzhumov for useful discussions.


\begin{thebibliography}{10}
\newcommand{\enquote}[1]{``#1''}
\expandafter\ifx\csname url\endcsname\relax
  \def\url#1{\texttt{#1}}\fi
\expandafter\ifx\csname urlprefix\endcsname\relax\def\urlprefix{URL }\fi
\providecommand{\eprint}[2][]{\url{#2}}

\bibitem{Smith00} D. R. Smith, W. J. Padilla, D. C. Vier, S.C. Nemat-Nasser, and S. Schultz, Phys. Rev. Lett. {\bf 84}, 4184--4187 (2000).

\bibitem{Shelby01} R. A. Shelby, D. R. Smith, and S. Schultz, Science {\bf 292}, 77--79 (2001).

\bibitem{Veselago68} V. G. Veselago Sov. Phys. Usp. {\bf 10}, 509--514 (1968).

\bibitem{Pendry99} J. B. Pendry, A. J. Holden, D. J. Robbins, and W. J. Stewart, IEEE Transactions on Microwave Theory and Techniques {\bf 47} 2075--2084 (1999).

\bibitem{cloak} D. Schurig, J. J. Mock, B. J. Justice, S. A. Cummer, J. B. Pendry, A. F. Starr, D. R. Smith, Science {\bf 314}, 977--980  (2006).

\bibitem{Pendry00} J. B. Pendry, Phys. Rev. Lett. {\bf 85} 18 3966--3969 (2000).

\bibitem{superlens} D. R. Smith, Science, {\bf 308}, 502 (2005).

\bibitem{blackbody} X. Liu,  T. Tyler, T. Starr, A. F. Starr, N. M. Jokerst, and W. J. Padilla, Phys. Rev. Lett. {\bf 107}, 045901 (2011).

\bibitem{MaM1} T. J. Yen, W. J. Padilla, N. Fang, D. C. Vier, D. R. Smith, J. B. Pendry, D. N. Basov, X. Zhang, Science {\bf 303} 1494--1496 (2004).

\bibitem{MaM2} S. Linden, C. Enkrich, M. Wegener, J. Zhou, T. Koschny, and C. M. Soukoulis, Science {\bf 306} 1351--1353 (2004)

\bibitem{MaM3} C. Enkrich, M. Wegener, S. Linden, S. Burger, L. Zschiedrich, F. Schmidt, J. F. Zhou, Th. Koschny, and C. M. Soukoulis, Phys. Rev. Lett. {\bf 95} 203901 (2005).

\bibitem{Magneto02} E. Shamonina, V. A. Kalinin, K. H. Ringhofer, and L. Solymara, J. Appl. Phys. {\bf 92} No.10 6252--6261 (2002).

\bibitem{Wiltshire07}   M. C. K Wiltshire, Phys. Status Solidi B-basic Solid State Phys. {\bf 224} 1227--1236 (2007).

\bibitem{RFM1} S. Tricarico F. Bilotti L. Vegni, IET Microwave Antennas Propag., {\bf 4} Iss.8 1026--1038 (2010).

\bibitem{RFIDM1} F. Paredes, G. Z. Gonzalez, J. Bonache, F. Martin, IEEE Trans. on Microwave Theory and Tech. {\bf 58} 1159--1166 (2010).

\bibitem{Wiltshire03}   M. C. K Wiltshire, J. V. Hajnal, J. B. Pendry, et al. Opt. Express  {\bf 11}  Issue:7 709--715 (2003).

\bibitem{Marques05} M. J. Freire, R. Marques, Appl. Phys. Lett. {\bf 86} iss:18 182505 (2005).

\bibitem{WPT1} Yaroslav Urzhumov and David R. Smith, Phys. Rev. B {\bf 83} 205114 (2011)

\bibitem{WPT2} B. Wang, K. H. Teo, T. Nishino, W. Yerazunis, J. Barnwell, and J. Zhang, Appl. Phys. Lett. {\bf 98} 254101 (2011)

\bibitem{Marques04} J. D. Baena, R. Marques, F. Medina, and J. Martel, Phys. Rev. B {\bf 69}, 014402 (2004).

\bibitem{mini07} Filiberto Bilotti, Alessandro Toscano, and Lucio Vegni, IEEE Trans. Antennas Propag. {\bf 55} NO. 8 (2007).

\bibitem{Super10} C. Kurter, J. Abrahams; S. M. Anlage, Appl. Phys. Lett. {\bf 96} Issue:25 253504  (2010)

\bibitem{Marques03} R. Marques, F. Mesa, J. Martel, and F. Medina, IEEE Trans. Antennas Propag. {\bf 51} Issue: 10 2572--2581 (2003).
    
\bibitem{Marques02} R. Marques, F. Medina and R. Rafii-El-Idrissi, Phys. Rev. B {\bf 65}, 144440, (2002)

\bibitem{Smith05} D. R. Smith, D. C. Vier,Th. Koschny, and C. M. Soukoulis, Phys. Rev. E {\bf 71} , 036617 (2005).
    
\bibitem{AM} See Supplemental Material at  (URL will be inserted by publisher) for the optical constant retrieval method used for the sprial metamaterials.

\end{thebibliography}
\end{document}